\documentclass{article}
\usepackage[preprint]{spconf}
\usepackage{amsmath,graphicx}

\usepackage{enumitem}
\setlist{nosep, leftmargin=14pt}

\usepackage{mwe} 
\usepackage{booktabs}
\usepackage{amssymb}
\usepackage{dsfont}
\usepackage{hyperref}

\title{Cinepro: Robust Training of Foundation Models for Cancer Detection in Prostate Ultrasound Cineloops}
\name{\begin{minipage}{\textwidth}
\begin{center}
   Mohamed Harmanani$^{1,5}$,
   Amoon Jamzad$^{1,5\star}$, \thanks{$\star$ denotes equal contribution.}
   Minh Nguyen Nhat To$^{2, 5\star}$, 
   Paul F.R. Wilson$^{1,5\star}$, \\
   \textit{Zhuoxin Guo}$^{1}$,
   \textit{Fahimeh Fooladgar}$^{2,5}$,
   \textit{Samira Sojoudi}$^{2}$,
   \textit{Mahdi Gilany}$^{1, 5}$,
   \textit{Silvia Chang}$^{2,3}$, \\
   \textit{Peter Black}$^{2,3}$,
   \textit{Michael Leveridge}$^{1,4}$,
   \textit{Robert Siemens}$^{1,4}$,
   \textit{Purang Abolmaesumi}$^{2\dagger}$,
   \textit{Parvin Mousavi}$^{1,5\dagger}$\thanks{$\dagger$ denotes co-senior authorship.}\thanks{© 2025 IEEE. Personal use of this material is permitted. Permission from IEEE must be obtained for all other uses, in any current or future media, including reprinting/republishing this material for advertising or promotional purposes, creating new collective works, for resale or redistribution to servers or lists, or reuse of any copyrighted component of this work in other works.}
   \end{center}
   \end{minipage}
}

\address{$^1$Queen's University, Kingston, Canada\\ 
$^2$University of British Columbia, Vancouver, Canada \\ 
$^3$Vancouver General Hospital, Vancouver, Canada\\
$^4$Kingston Health Sciences Center, Kingston, Canada\\ 
$^5$Vector Institute, Toronto, Canada
}

\begin{document}
\maketitle
\begin{abstract}
Prostate cancer (PCa) detection using deep learning (DL) models has shown potential for enhancing real-time guidance during biopsies. However, prostate ultrasound images lack pixel-level cancer annotations, introducing label noise. Current approaches often focus on limited regions of interest (ROIs), disregarding anatomical context necessary for accurate diagnosis. Foundation models can overcome this limitation by analyzing entire images to capture global spatial relationships; however, they still encounter challenges stemming from the weak labels associated with coarse pathology annotations in ultrasound data. We introduce Cinepro, a novel framework that strengthens foundation models' ability to localize PCa in ultrasound cineloops. 
Cinepro adapts robust training by integrating the proportion of cancer tissue reported by pathology in a biopsy core into its loss function to address label noise, providing a more nuanced supervision. Additionally, it leverages temporal data across multiple frames to apply robust augmentations, enhancing the model’s ability to learn stable cancer-related features. 
Cinepro demonstrates superior performance on a multi-center prostate ultrasound dataset, achieving an AUROC of 77.1\% and a balanced accuracy of 83.8\%, surpassing current benchmarks. These findings underscore Cinepro's promise in advancing foundation models for weakly labeled ultrasound data.

\end{abstract}
\section{Introduction}
\label{sec:intro}
Early detection of prostate cancer (PCa) is crucial to improve patient outcomes and reduce the risk of mortality. PCa detection is performed using systematic biopsy under the guidance of trans-rectal ultrasound (TRUS), by sampling 8-12 predefined anatomical locations of the prostate, followed by histopathological analysis of the cores. The standard of care is starting to shift to transperineal ultrasound guided biopsy. In general, systematic biopsy has a low sensitivity, reported to be between 40-50\%~\cite{ahmed2017diagnostic}. This limitation arises because the sampling is inherently blind to variations in tissue appearance and may miss cancer outside the target locations.

Given this challenge, there is a growing interest in leveraging deep learning (DL) to enhance biopsy accuracy. DL models can analyze ultrasound or other imaging modalities in real time~\cite{huang2023transfer, jiang2024microsegnet}, providing the clinician with precise tissue characterization to better guide biopsy procedures. This has the potential to improve detection rates, reduce the number of unnecessary biopsies, and lead to earlier, more effective treatment.
\begin{figure*}[h!]
    \centering
    \includegraphics[width=2.03\columnwidth]{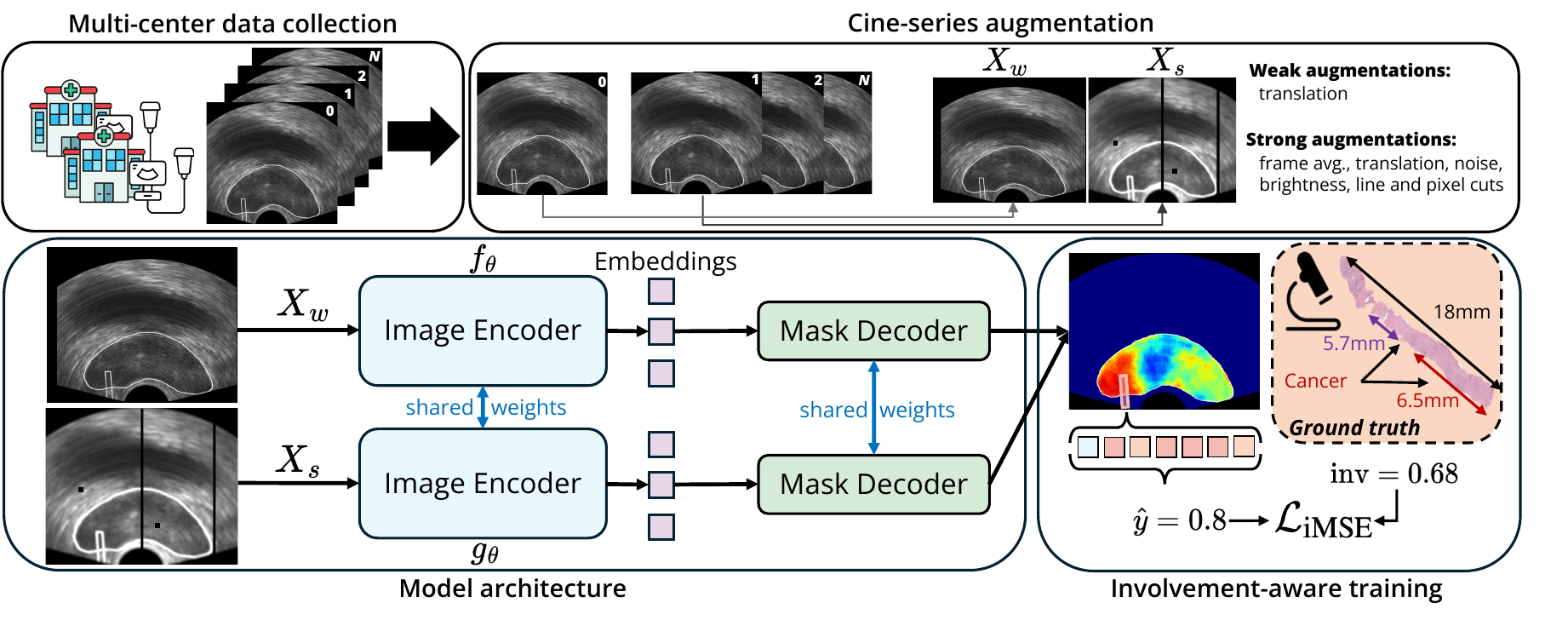}
    \caption{Robust training of foundation models on a multi-center dataset for PCa detection in conventional ultrasound. We train MedSAM to predict the involvement reported by pathology, and add a cine-series augmentation strategy to increase robustness.}
    \label{fig:cinepro_method}
\end{figure*}
Capturing the broader anatomical context is crucial for accurate PCa detection in ultrasound, as subtle textural patterns often emerge only in relation to surrounding tissue. However, existing approaches remain limited in their ability to incorporate this context effectively. For instance, ROI-based methods focus on specific regions along the biopsy track~\cite{wilson2024toward}, limiting their ability to capture information beyond these small regions. Multiple-instance learning (MIL) methods~\cite{bulten2022artificial} improve on this by analyzing bags of features that represent local patches, but still fall short of leveraging the entire image. Hence, both ROI and MIL methods are constrained in their ability to account for the prostate's global structure and surrounding regions. In contrast, whole-image methods~\cite{huang2023transfer, wilson2024prostnfound} have shown promising results for PCa detection, demonstrating the value of incorporating full-image information.

Foundation models, such as MedSAM~\cite{ma2024segment}, further enhance these whole-image methods by leveraging their extensive pre-training on large, diverse datasets. This allows them to capture global contextual information across the entire image. However, adapting foundation models for PCa detection presents unique challenges. While MedSAM excels at segmenting organs with clearly defined boundaries, cancer detection requires the model to identify subtle, fine-grained textural variations within the same tissue that are often difficult to discern~\cite{rohrbach2018high}. 
This challenge is exacerbated by the lack of dense pixel-wise annotations of cancer in ultrasound data. Instead, foundation models must rely on weak labels produced by pathology, which provide only coarse information about the presence of cancer rather than precise delineations. This calls for the development of robust strategies to effectively train on imperfect data while still taking advantage of the foundation model's pre-trained knowledge.

To address these challenges, we propose Cinepro, a novel strategy for training robust foundation models on weakly labeled prostate ultrasound cineloops. Our contributions include: (i) adapting segmentation foundation models to cancer classification on a multi-center dataset of prostate cancer from conventional ultrasound; 
(ii) using cancer involvement (proportion of malignant tissue) reported by pathology as part of the loss function in order to contend with label noise by providing the model with a stronger form of supervision; and (iii) design of a new weak-strong data augmentation strategy using multiple cine-series frames to train the model on challenging yet realistic variations of the tissue over time, further enhancing its robustness. Cinepro surpasses common benchmarks and previous literature, with an AUROC of 77.1\%.

\begin{table}[ht!]
    \centering
    \caption{Summary statistics of our data}
    \begin{tabular}{lccccccc}
    \toprule
        \bf Ctr. & \bf Cases & \bf Benign & \bf GS7 
        & \bf GS8 & \bf GS9 & \bf GS10\\
    \midrule
    VGH & 131 & 680 & 120 & 52 & 24 & 7\\
    KHSC& 180 & 1110 & 166 & 35 & 32 & 5 \\
    \midrule
    \bf Total & 311 & 1790 & 286 & 87 & 56 & 12 \\
    \bottomrule
    \end{tabular}
    \label{tab:data}
\end{table}

\section{Materials and Methods}
\label{sec:methods}

\begin{table*}[t]
    \centering
    \caption{Comparing the performance of Cinepro to prior baselines and augmentation methods for PCa detection. Metrics are averaged across folds, and the standard deviation is reported.}
    \begin{tabular}{lccccccc}
        \toprule
         \bf Method & \bf AUROC & {$\underset{\text{inv} >0.35}{\textbf{AUROC}}$} & \bf Bal. Acc. & \bf Sens.@20Spe & \bf Sens.@40Spe  & \bf Sens.@60Spe \\
         \midrule

         ROI-based iLR~\cite{to2022coarse} & 67.0 $\pm$ 0.7 & 70.2 $\pm$ 0.7& 63.7 $\pm$ 0.7 & 93.6 $\pm$ 1.0 & 83.1 $\pm$ 1.5 & 67.0 $\pm$ 1.5\\ 
         UNet +MaskCE~\cite{wilson2024prostnfound} & 67.5 $\pm$ 1.5 & 71.4 $\pm$ 1.6 & 64.1 $\pm$ 1.2 & 92.9 $\pm$ 0.7 & 83.9 $\pm$ 0.9 & 65.5 $\pm$ 3.8\\
         UNet +iMSE & 68.8 $\pm$ 1.5 & 73.5 $\pm$ 1.8 & 65.6 $\pm$ 1.5 & 92.0 $\pm$ 1.1 & \bf 86.9 $\pm$ 0.8 & 70.4 $\pm$ 3.4\\
         \midrule
         SAM +MaskCE~\cite{wilson2024prostnfound} & 71.2 $\pm$ 3.4 & 77.0 $\pm$ 5.6 & 68.4 $\pm$ 3.7 & 90.6 $\pm$ 2.8 & 80.6 $\pm$ 1.6 & 68.4 $\pm$ 6.6\\
         SAM +iMSE & 71.6 $\pm$ 4.4 & 78.7 $\pm$ 3.2 & 67.5 $\pm$ 3.5 & 90.8 $\pm$ 4.7 & 82.9 $\pm$ 6.2 & 71.4 $\pm$ 6.3\\
         MedSAM +MaskCE~\cite{wilson2024prostnfound} & 73.9 $\pm$ 3.2 & 78.3 $\pm$ 3.2 & 68.9 $\pm$ 1.9 & 94.2 $\pm$ 6.0 & 81.9 $\pm$ 3.7 & 71.7 $\pm$ 5.4\\
         MedSAM +iMSE & 75.2 $\pm$ 2.6 & 81.5 $\pm$ 2.1 & 69.3 $\pm$ 2.3 & 93.3 $\pm$ 1.3 & 85.7 $\pm$ 4.3 & 75.6 $\pm$ 4.0\\
         \midrule
         SAM +iMSE+WS-Augs. & 75.6 $\pm$ 0.7 & 81.6 $\pm$ 2.3 & 70.1 $\pm$ 0.8 & 93.7 $\pm$ 2.0 & 84.1 $\pm$ 2.3 & 75.3 $\pm$ 0.6\\
         MedSAM +iMSE+WS-Augs. & 76.1 $\pm$ 0.6 & 82.4 $\pm$ 0.6 & 70.9 $\pm$ 1.2 & 92.9 $\pm$ 1.2 & 85.9 $\pm$ 1.1 & 77.5 $\pm$ 1.6\\
         \midrule
         Cinepro (ours) & \bf 77.1 $\pm$ 2.0 & \bf 83.8 $\pm$ 1.8 & \bf 71.9 $\pm$ 1.6 & \bf 94.6 $\pm$ 1.0 & 85.7 $\pm$ 3.4 & \bf 78.1 $\pm$ 4.0\\
        \bottomrule
    \end{tabular}
    \label{tab:tab1_results}
\end{table*}

\subsection{Data}
\textbf{Acquisition:} We use private data from 311 patients in two clinical centers: Kingston Health Sciences Center (KHSC) in Ontario and Vancouver General Hospital (VGH) in British Columbia, Canada (Table~\ref{tab:data}). The study was approved by the institutional health research ethics boards and all patients provided informed verbal and written consent to participate.
Raw radio-frequency (RF) ultrasound data are acquired using BK3500 ultrasound equipped with an E10C4 endocavity transducer. At each biopsy location, 200 consecutive frames of RF data are acquired by holding the transducer still for 3-5 seconds. An average of 10 to 12 biopsy cores are extracted from each patient, resulting in a cineloop of 200 ultrasound frames of size $1960\times 282$ for each core. 

\textbf{Preprocessing:} We convert the RF data to B-Mode images and manually select the biopsy needle region where the tissue cores are extracted. We match the data with the corresponding histopathology report and label the pixels in the needle region as cancer or benign, accordingly. The images are then resized to $1024\times 1024$. Finally, we exclude any data containing motion and acoustic shadowing artifacts.

\subsection{Involvement-aware training}
\textbf{Model architecture:} We use a pre-trained foundation model with the same weights and architecture as MedSAM~\cite{ma2024segment}. Our overall methodology and training pipeline is shown in Figure~\ref{fig:cinepro_method}. The architecture consists of a vision transformer (ViT) image encoder, with approximately $90M$ parameters, and a mask decoder with approximately $6M$ parameters. The model is trained to segment cancer in the US image, and is given the involvement in the needle region as a ground-truth. The encoder then generates a $256\times256\times64$ embedding, which the mask decoder uses to produce a prediction matrix of size $256\times256$. To make a final prediction, we select the pixels in the needle region and average their activation values, as computed by the model, and use this score as a pseudo-probability of cancer. 

\textbf{Loss function:} Let $\mathcal{R}$ be the set of pixels in the region formed by the intersection of the needle region $N$ and the prostate mask $P$ given by $\mathds{1}_{N[i,j]\ne0 \wedge P[i,j]\ne0}$ for $i,j \in [1,1024]$. We define the involvement-aware loss function to minimize the distance between the average intensity of the model's predictions within the region of interest $\mathcal{R}$ and the reported cancer involvement of the core:
\begin{align*}
    \mathcal{L}_{\text{iMSE}}(\hat{Y}, \text{inv}) &= \frac{1}{|\mathcal{R}|}\sum_{i=0}^{1024}\sum_{j=0}^{1024} (\hat{Y}[i,j] \cdot\mathds{1}_{\hat{Y}[i,j] \in \mathcal{R}} - \text{inv})^2 .
\end{align*}
By training the foundation model to localize cancer optimized by the involvement, we aim to mitigate the impact of coarse labeling by allowing the model to learn some information regarding the spread of cancer in the core.

\subsection{Cine-series augmentation for robust training}
To make the foundation model robust to weak labeling, we develop a training strategy using cine-series data augmentations specifically tailored to ultrasound data. Loosely inspired by the literature on weak-strong augmentations in semi-supervised learning~\cite{sohn2020fixmatch, yang2023revisiting}, the proposed strategy enhances the model's ability to generalize across time-dependent variations, ensuring robustness in scenarios where the characteristics of the tissue change over time. In this approach, we take the initial frame in the US sequence and apply a ``weak" augmentation, resulting in $X_w$. We experimented with various types of data augmentations and found that simple random translations were sufficient to improve performance considerably. We then take the remaining 199 frames and compute their average, before applying another series of augmentations such as brightness jitter, speckle noise, salt and pepper noise, random translations, and random line and pixel cuts. We refer to this process as ``strong" augmentation, and the resulting image is denoted as $X_s$. We then train two models $f_\theta$ and $g_\theta$ with shared weights to localize cancer in $X_w$ and $X_s$. The apply a confidence threshold, $\tau$, to the output $f_\theta(X_w)$ to remove lower intensity pixels and compute the final output of the model using a weighted average:
\begin{align*}
    \hat{Y} &= {\gamma_w [f_\theta(X_w)\cdot\mathds{1}_{f_\theta(X_w)>\tau}] + \gamma_s g_\theta(X_s)} ,
\end{align*}
where $\tau, \gamma_w, \gamma_s$ are all tuneable hyperparameters, and $\gamma_w + \gamma_s = 1$. We then compute the involvement-aware loss given by $\mathcal{L}_{\text{iMSE}}(\hat{Y}, \text{inv})$ and update the weights. 

Our new training strategy, together with involvement-aware loss, further strengthens training by maintaining a strong correlation between actual cancer involvement and predicted regions, and reducing the impact of noise and irrelevant variations in the data over time.

\begin{figure*}[h!]
    \centering
    \includegraphics[width=2\columnwidth]{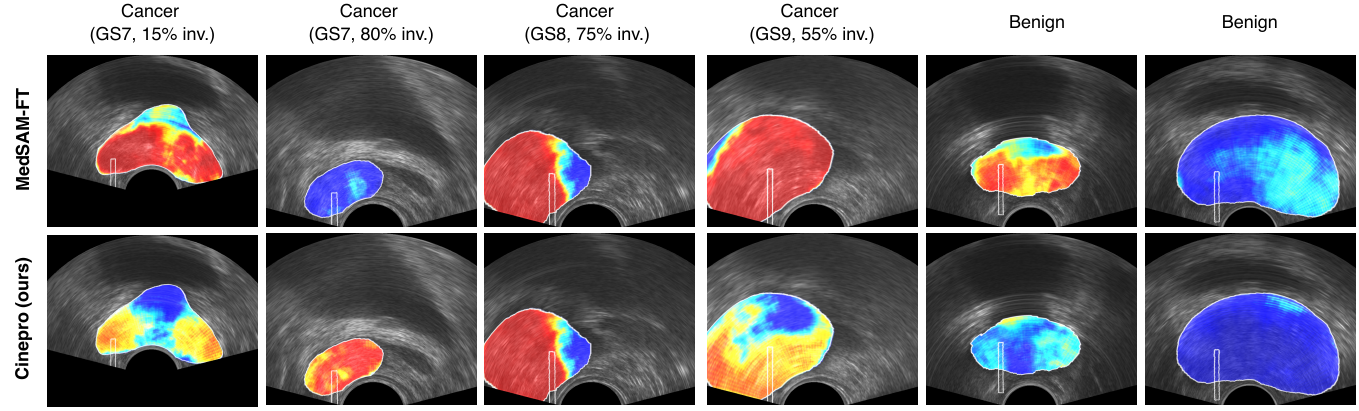}
    \caption{Qualitative comparison of Cinepro with our finetuned MedSAM baseline.}
    \label{fig:heatmaps}
\end{figure*}

\subsection{Experiments}
We compare Cinepro to several baselines for PCa detection, spanning architectures, loss functions, and augmentation strategies.  
Our first baseline is involvement-based label refinement (iLR)~\cite{to2022coarse}, a noise-tolerant ROI-based method that also uses the involvement to train a PCa classifier. 
We also train UNet, SAM, and MedSAM with both a masked cross-entropy (MaskCE) objective~\cite{wilson2024prostnfound}, and our new iMSE loss function. For augmentation, we evaluate naive random translation, weak-strong augmentations, and our cine-series approach. 
All models are trained and evaluated using a 5-fold cross-validation scheme. A detailed breakdown of hyperparameter tuning and computational efficiency is provided in our GitHub repository: \href{https://github.com/mharmanani/cinepro}{https://github.com/mharmanani/cinepro}.

\section{Results}
\label{sec:results}
\textbf{Quantitative results:} Our experimental results are shown in Table~\ref{tab:tab1_results}. We observe that whole-image methods outperform the iLR baseline considerably, highlighting their advantage over ROI methods. Moreover, foundation models exceed other baselines in performance, with MedSAM and SAM consistently outperforming UNet across most metrics. 

Our proposed loss function, iMSE, yields better overall performance than MaskCE across most metrics. SAM+iMSE achieves a slightly higher AUROC (+0.4\%) than SAM with MaskCE, but has a lower balanced accuracy (-1.1\%). Notably, iMSE leads to superior cancer detection, as evidenced by higher sensitivity across various thresholds, but comes at the cost of increased false positives. In contrast, when used with MedSAM, iMSE outperforms MaskCE in AUROC by a wider margin while also producing better sensitivities across all thresholds. MedSAM's pre-training has likely amplified iMSE's advantage over MaskCE (+1.3\% AUROC, +3.2\% Bal.Acc.) by enabling more precise localization of cancer, and thus, better correlation with cancer involvement in the core. 
The impact of the weak-strong augmentation strategy (without cine-series) is also notable, with the dual-augmented versions of SAM and MedSAM outperforming their base counterparts (with translation) by 6\% and 1.1\% respectively. The resulting performance gain is smaller for MedSAM, suggesting that the augmentation strategy is more suited for generalist models.

Finally, our proposed method for robust training of foundation models achieves superior performance across most metrics, with AUROC and balanced accuracy improving by +1.9\% and +2.3\% over the strongest baseline. iMSE plays a pivotal role by aligning predictions with cancer involvement, enabling the model to capture subtle, cancer-relevant patterns more effectively. Combined with cine-series augmentation, iMSE reinforces temporal consistency, helping the model focus on stable, disease-related features and reducing sensitivity to noise or irrelevant variations. This combination results in more reliable and accurate predictions, with sensitivity improving to 78.1 at 60\% specificity (+2.5\%), further demonstrating the model’s advantage in cancer detection and resilience to false positives.

\textbf{Qualitative results:} Figure~\ref{fig:heatmaps} shows prediction heatmaps using both Cinepro and the fine-tuned MedSAM approaches described earlier. We show four cancer examples with different Gleason scores and cancer involvement, and 2 benign examples. We observe that the activations of the output masks generated by Cinepro are less strong in low-involvement cores, thus providing a more accurate form of tissue characterization that matches the involvement, than the baseline. This is especially true for involvement values in the range of 15-70\%, with Cinepro producing average predictions that are close to the true distribution of cancer. For high involvement cores, both methods are comparable and perform well, but Cinepro outperforms fine-tuned MedSAM by achieving a higher true positive rate. Finally, the performance of the two models on benign cores is also comparable, but MedSAM produces more false positives in the needle region. 

\section{Conclusion}
\label{sec:conclusion}
We proposed Cinepro, a novel adaptation of segmentation-based foundation models to prostate cancer classification from ultrasound. We further introduced a robust framework for fine-tuning foundation models on weakly labeled prostate ultrasound data, using a involvement-driven loss function to increase the model's robustness to noise through stronger supervision. We subsequently introduced a novel cine-series data augmentation strategy that significantly improves model reliability, reducing noise and increasing sensitivity, particularly in high-specificity settings. These findings highlight the potential of combining temporal learning with tailored loss functions and medical-specific models to advance automated cancer detection and improve early diagnosis. Future work should focus on integrating the proposed method in a larger study to further improve generalization and robustness.

\section{Compliance with ethical standards}
\label{sec:ethics}

This study was approved by the Institutional Research Ethics Board at both sites, and patients provided informed consent to participate.

\section{Acknowledgements}
\label{sec:acknowledgements}
This work was supported by the Natural Sciences and Engineering Research Council of Canada (NSERC), and the Canadian Institutes of Health Research (CIHR). Parvin Mousavi is supported by the CIFAR AI Chair and the Vector Institute. None of the other authors have potential conflicts of interest to disclose.

\bibliographystyle{IEEEbib}

\end{document}